\documentclass[12pt,a4paper]{article}              
\usepackage{amsmath,graphics,graphicx,amssymb,anysize,natbib,color,setspace,pifont}

\renewcommand{\refname}{}

\marginsize{25mm}{25mm}{20mm}{20mm}

\begin{document}

\begin{onecolumn}
\section*{\textsf{Quasicrystal nucleation in an intermetallic glass-former}}
Wolfgang Hornfeck, Raphael Kobold$^{*}$, Matthias Kolbe, Dieter Herlach \\ \\
Institut f\"{u}r Materialphysik im Weltraum, Deutsches Zentrum f\"{u}r Luft- und Raumfahrt (DLR), 51170 K\"{o}ln, Germany. ($^{*}$ raphael.kobold@dlr.de).
%

\vfill

\textsf{The discovery of quasicrystals 30~years ago challenged our understanding of order at the atomic scale.\cite{Shechtman} While quasicrystals possess long-range orientational order they lack translation periodicity.\cite{SteurerBuch} Structurally complex, yet crystalline intermetallics\cite{ComplexMetallicAlloys2} and (bulk) metallic glasses\cite{Greer1,Greer2} represent competing states of condensed matter among metallic phases, including peculiarities like the $q$-glass.\cite{qglass} Considerable progress has been made in their structure elucidation and in identifying factors governing their formation\cite{QCrecent1,QCrecent2,QCrecent3} -- comparatively less is known about their interrelation. Moreover, studies bridging the spatial scales from atoms to the macro\-scale are scarce. Here we report on the homogeneous nucleation of a single quasicrystalline seed of decagonal symmetry and its continuous growth into a tenfold twinned dendritic microstructure. Electrostatic levitation was applied to undercool melts of glass-forming NiZr, observing \emph{single} crystallization events with a high-speed camera -- with a statistical evaluation of 200~\emph{consecutive} thermal cycles suggesting homogeneous nucleation. The twinned dendritic microstructure, apparent in electron backscatter diffraction maps, results from the symmetry breaking of an essentially 2D decagonal quasicrystalline cluster. Conserving its long-range orientational order, our distortion-free twin model merges a common structure type for binary intermetallic compounds, with interatomic distances of alike atoms scaled by the golden ratio, and spiral growth resembling phyllo\-taxis. NiZr represents a missing link connecting quasicrystals and multiple twinned structures sheding light on intermediate states of order between glasses, crystals and their twins, and quasicrystals.}

\newpage

A first account relating a dendritic growth morphology of a crystal to its structure is Kepler{'}s 1611 treatise \emph{On the Six-Cornered Snowflake}\cite{Kepler} studying the closest sphere packings in 2D and 3D, distinguishing the (face-centered) cubic (\textit{fcc}) from the hexa\-gonal one (\textit{hcp}). Explaining most of the crystal structures of metals, an atom is surrounded by 12~nearest neighbours in a cuboctahedral (\textit{fcc}) or anti-cubocta\-hedral (\textit{hcp}) shell. Yet another, and locally more dense, high-symmetry coordination of 12~atoms around a central one is achieved by the icosahedron, one of the Platonic solids known since antiquity. However, its rotational symmetry is not compatible with lattice translations. For this reason, the icosahedron features in pioneering theories about the structure of liquids as well as in Frank{'}s explanation\cite{Frank} of Turnbull{'}s empirical studies\cite{Turnbull} on the \emph{undercoolability of metallic melts}. Following Frank{'}s argument, high undercoolings are possible when the local structures of the solid and the melt differ from each other. Then, a structural reorganization prior to solidification is required, acting as a barrier for nucleation.\cite{HerlachISRO,KeltonISRO} Non-crystallographic, yet dense packing of spheres into icosahedral shells were envisaged by Mackay.\cite{MackayICO} Mackay was also the first to explore a {'}\emph{penta\-gonal snowflake}{'},\cite{MackaySnowflake} i.e.~a crystallographic realization of Penrose{'}s tiling of {'}forbidden{'} pentagonal symmetry, and to predict its diffraction pattern\cite{MackayPenrose} prior to Shechtman{'}s paradigm-shifting discovery of quasicrystals.\cite{Shechtman} A fierce controversy about their existence, in opposition to traditional explanations based on multiple twinning put forward by Pauling, was eventually settled by awarding Shechtman the 2011 Nobel prize for chemistry. In a \emph{l{'}esprit de l{'}escalier} we like to discuss a special nucleation and growth mechanism intimately connecting a quasicrystalline seed with a multiple twinned crystal structure in what may called: \emph{A decagonal snowflake}.

Binary Ni--Zr alloys form complex crystal structures and, upon rapid solidification, bulk metallic glasses. In the chemically related ternary system with Ti~an icosahedral quasicrystal exists.\cite{KeltonTi} This makes Ni--Zr a perfect system to study the structural interplay between complex crystals, quasicrystals and glasses. Ni$_{x}$Zr$_{1-x}$ vitrifies into a glass within a range of $0.2 < x < 0.7$ comprising several eutectics. On the contrary, the line compound NiZr is a congruently melting glass-former. This facilitates systematic studies of dendritic growth in a chemically simple, compositionally well-defined metallic alloy. Non-equilibrium solidification experiments of NiZr were performed in an undercooling regime bounded by its melting and glass transition ($T_{\mathrm{m}} = 1533$\,K, $T_{\mathrm{g}} = 730$\,K). 

Electrostatic levitation (ESL) is a state-of-the-art method for containerless processing of materials in an ultra-high vacuum thereby maintaining high purity conditions. The sample is levitated and actively positioned by electrostatic fields and independently heated with an infrared laser. A typical processing cycle includes the melting and (under-)\-cooling of the sample, until spontaneous crystallization occurs at the nucleation temperature $T_{\mathrm{n}}$, accompanied by a local release of latent heat. The maximum undercoolings achieved, $\Delta T = T_{\mathrm{m}}-T_{\mathrm{n}} = 300(5)$\,K, vary only in the range of a few Kelvin. The intersection of the moving solidification front with the sample{'}s surface, observed with a high-speed camera (HSC), reproducibly shows a decagonal shape growing with an average velocity of $\langle v \rangle = 0.51(4)$\,m/s (Fig.~\ref{hsc}a). 

The microstructures of as-solidified samples were analyzed by using optical polarization and scanning electron microscopy (SEM), including energy dispersive X-ray spectrometry (EDS) for chemical analysis and electron backscatter diffraction (EBSD). EBSD is a powerful method to determine and visualize the crystallographic orientation of intergrown grains, either via a 2D~false-colour map or an inverse pole figure plot (IPF), i.e.~a stereographic projection of observed crystallographic directions.\cite{RappazEBSD}

Solidified, spherical samples were overcasted with numerous fine geodesics resembling the mesh of lattitudes and longitudes on a globe. \emph{Ten} longitudes meet at two poles marking the onset point of solidification and its antipode, allowing for an oriented embedding. Series of cross-sections, showing the 3D-evolution of microstructural features, confirm the alignment of the putative growth direction with the tenfold axis.
The main feature is the presence of \emph{ten} major grains separated by coherent boundaries traversing the sample with an angular inclination of $\sim 36^{\circ}$ (Fig.~\ref{ebsd}a,b).

While the tenfold pattern of major grains, originating in perfect registry from a single center, governs the impression, each grain exhibits a (coarsened) dendritic fine structure, resembling the \emph{feathery grains} of directed solidification experiments.\cite{RappazAlu} In particular, both coherent and incoherent, straigth and wavy grain boundaries, are simultaneously present, most often in an alternating arrangement (Fig.~\ref{ebsd}c,d). This appearance is caused by a two-fold twinning, by which the stem of every dendrite is constituted by a large-angle grain boundary with a $36^{\circ}$ twin orientational relationship (Fig.~\ref{ebsd}b, Fig.~\ref{lunker}). Thus, the microstructure exhibits a spatial hierarchy of structural organization. Twinned dendrites are visible up to the third order (Fig.~\ref{ebsd}c). Additional deformation twinning is clearly distinguished by its all-parallel straight twin boundaries. 

Similar microstructures were found in the vicinity of NiZr ($\pm 2\,\mathrm{at.\%}$), for ternary substitution variants (Ni,Cu)Zr, Ni(Zr,Hf) with about the same amount of minority component (Supplemental information), as well as for CrB-type AlZr.

NiZr crystallizes in a CrB-type structure comprising eight atoms in a $C$-centered, orthorhombic unit-cell of space group $Cmcm$, $a = 326.8(8)$\,pm, $b = 993.7(4)$\,pm, $c = 410.1(5)$\,pm. Both Ni and Zr occupy the Wyckoff site $4\,c$ $(0,y,1/4)$ with $y_{\mathrm{Ni}} = 0.0817(17)$ and $y_{\mathrm{Zr}} = 0.3609(8)$.\cite{Kirk} The axial ratio $a/b$ of the NiZr structure is such, that the $(\pm1\,1\,0)$ pair of net planes encloses an angle of almost the \emph{tenth} of a full circle: $\varphi = 2\,\tan^{-1}(a/b) \sim 36.4^\circ$.

This simple but special geometrical relation is a prerequisite for tenfold microtwins (TMTs). TMTs were observed in splat-cooled NiZr using high-resolution transmission electron microscopy (HRTEM).\cite{KuoNiZr} They nucleate upon heating from an amorphous matrix of melt-spun NiZr and (Ni$_{0.9}$Cu$_{0.1}$)Zr, notably before any other phase formation is observed.\cite{Foga} \emph{Dendritic} TMTs also form in thin films of metastable Cr$_{3}$C$_{2-x}$.\cite{Bouzy} In fact, already the CrB-type results from a \emph{periodic unit-cell twinning} of the  \textit{fcc}-type.\cite{Parthe,Andersson} In any case the lateral extension of TMTs was restricted to only a few micrometers, whereas for ESL-processed NiZr undistorted twin boundaries extend up to the samples diameter. This extraordinary fact requires an explanation! 

Based on the aforementioned specific geometrical properties of CrB-type NiZr we propose an atomistic model (Fig.~\ref{structure}a), reconstructing the central part of a TMT, thereby extending prior (1985) HRTEM-results of Kuo~\textit{et al.}\cite{KuoNiZr} to the atomic scale. The key idea concerns the \emph{equalization} of two pairs of (projected) short $S{'}=\sqrt{a^{2}+(L-S)^{2}}\big/2\stackrel{!}{=}S$ and long $L{'}=\sqrt{a^{2}+(b-2L)^{2}}\big/2\stackrel{!}{=}L$ interatomic distances of the NiZr structure (Fig.~\ref{structure}c) within neighbouring twin domains and across their boundaries (Fig.~\ref{structure}a).

This leads to a unique choice of structural parameters for NiZr: $b/a = \sqrt{5+2\sqrt{5}} \sim 3.078$, $S/a = \sqrt{(5-\sqrt{5})/10} \sim 0.526$, $L/a = \tau\,S/a \sim 0.851$, with $\tau \sim 1.618$ the \emph{golden ratio}, including ideal atomic coordinates for Ni, $y_{\mathrm{Ni}}/b = \tau^{-2}/(2\sqrt{5}) \sim 0.085$, and Zr, $y_{\mathrm{Zr}}/b = \tau/(2\sqrt{5}) \sim 0.362$. Contrary to common views of twinning, our model includes the metrical relations imposed by the Bravais lattice \emph{and} the spatial distribution of atoms within the NiZr unit-cell. Quite remarkably, the structure in the bulk is \emph{identical} to the structure \emph{across} the twin boundary! This exceptional feature of a \emph{distortion-free}, energetically advantageous twinning explains the dominant occurrence of twin boundaries as a macroscopic growth feature. 

A specific shift of $\sigma/a = \pm\,1/2 \Leftrightarrow \sigma/d = \pm\,(2\,\sqrt{1+(b/a)^2})^{-1} = \pm\,(4\,\tau)^{-1} \sim \pm\,0.155 \sim \pm\,1/6$ along the $\langle 110 \rangle$-directions of common twin planes (with $d = 2\,\tau\,a$ the length of a unit-cell diagonal $\boldsymbol{d} = \boldsymbol{a} + \boldsymbol{b}$) accompanies our twin mechanism. A \emph{common} choice of sign for all $\langle 110 \rangle$-shifts yields a \emph{chiral} arrangement of alike atoms into spirals forming alternating shells of a facetted \emph{cyclic intergrowth}\cite{Andersson} structure. Thus, an ideal twin center may exist in two mirror-symmetric forms.

The innermost core of the twin model, representing the putative frozen-in seed of nucleation, shows a $1\colon\tau\colon\tau^{2}$ scaling for three successive shells of atoms (Fig.~\ref{structure}a). The transition from this {'}quasicrystalline{'} cluster to the TMT-crystal is marked by a loss of higher-order scalings (Fig.~\ref{structure}a).
The twin model thus exhibits some characteristics of axial quasicrystals with decagonal symmetry.\cite{Decagonal} Its 2D Fourier map shows distinctive features of decagonal quasicrystals beyond the mere radial symmetry expected for a tenfold twin (Fig.~\ref{structure}b). These dual space features indicate a long-range decagonal orientational order in the real structure. Indeed, vertex-defective pentagons occur in all possible rotated and reflected orientations throughout the twin model. Notably, $\angle(S,S{'}) = 108^{\circ}$ and $\angle(L,L{'}) = 144^{\circ}$ correspond to the exterior angle of a regular pentagon and decagon, respectively.

Spiral growth in the $\boldsymbol{ab}$-plane can be easily extended into a microscopic pattern maintaining the symmetry of the nucleus. Upon dendrite formation a highly correlated, cooperative growth sets in, conserving the twin boundaries into the dendrite stems, yielding the geodetic pattern of surface twin faults. Along the $\boldsymbol{c}$-direction growth occurs by extending the innermost Zr-centered pentagonal antiprism of Ni-atoms, first into an (\emph{compressed}) icosahedron with two Zr-atoms at its apices and eventually into a pentagonal antiprismatic column, a structural motif well-known in many intermetallics. Thus, the in-plane spiral growth mechanism is amended with an out-of-plane columnar one, both of which feature preferred nucleation sites, favouring the attachment of atoms either in re-entrant kinks between neighbouring twin domains or along the staggered packing of pentagons into antiprisms (Fig.~\ref{structure}d). A compression of the central icosahedron breaks its icosahedral symmetry, with only one out of six fivefold symmetry axes remaining for columnar growth, favouring a $(2+1)$-dimensional growth model. A single pentagonal antiprismatic seed, once formed, appears sufficient to ensure undisturbed growth into a macroscopic twin (Fig.~\ref{structure}e). On the contrary, a recently reported twinning mechanism based on \emph{ideal} icosahedra, in which dopants trigger the formation of a quasicrystalline seed, enhancing the heterogeneous nucleation of a competing \textit{fcc}-type phase, improves the sample{'}s grain-refinement.\cite{RappazAlZnCr,RappazAuIr}

Specifically we propose that the compressed icosahedron indeed \emph{is} the primordial homogeneous nucleus, originating in the melt \emph{only} at high undercoolings. The case for \emph{homogeneous} nucleation is made by a series of Poisson-distributed nucleation events and its statistical analysis. This allows to determine thermodynamical and kinetic parameters,\cite{Klein} such as the critical free enthalpy of nucleation, $\Delta G^{*} = 59.717\,k_\mathrm{B} T_{\mathrm{n}}$, the solid-liquid (dimensionless) interfacial energies $\alpha = 0.7057$ and $\sigma = 0.1346(7)\,\mathrm{J}/\mathrm{m}^{2}$, as well as the pre-exponential factor $K_{V} = 6.38 \times 10^{34} \mathrm{m}^{-3}\mathrm{s}^{-1}$ of the nucleation rate, all matching the expected values for similar metallic systems.\cite{Turnbull,KeltonISRO,Klein} (Fig.~\ref{nucleation}; Supplemental information). 
Now, homogeneous nucleation of an icosahedral seed in deeply undercooled NiZr is plausible for several reasons: (i) the large undercoolings of small variation; (ii) the skew distribution of nucleation events with a steep decrease at higher relative undercoolings $\Delta T/T_{\mathrm{m}}$; (iii) the value of $\{K_{V}\} \sim 10^{34}$ compared to Turnbull{'}s estimates of $\sim 10^{25}$ for heterogeneous and $\sim 10^{39}$ for homogeneous nucleation in pure metals;\cite{Turnbull} and (iv) the value of $\alpha$ closely matching the one characterizing polytetrahedral (icosahedral) structures.\cite{KeltonISRO,Klein} Furthermore, an estimation of the critical radius $r^{*} = 1.3$\,nm of classical nucleation theory compares reasonably well with the radius $r/a = \tau^3 S \sim 0.7$\,nm for the quasicrystalline core of the twin model.

Generalizing the aforementioned geometrical considerations the 132~known binary CrB-type structures (Pearson{'}s Crystal Data 2012/13, ASM International) can be classified according to their axial ratios $b/a$ and $c/a$. In a twin description, \emph{all} structures exhibit a compressed central icosahedron, however with the geometrical flexibility and potential to form $n$-fold twins for $7 < n < 12$ (Fig.~\ref{icocrb}).

NiZr offers a unique insight into the formation of a highly-refined dendritic microstructure in a glass-forming alloy, based on the combination of fundamental crystallography with advanced yet generally applicable processing and characterization techniques. We propose a general geometric mechanism bridging the scales from a homogeneously nucleated quasicrystalline seed via twinning-induced growth to a macroscopic solid. In a most comprehensive view, our geometric ideas seems suited to model the quasicrystal--crystal transition for axial quasicrystals with even-fold rotational symmetry ($n = 8, 10, 12,\ldots$), in the generic case of a dense, binary sphere packing, with potential applications in materials design and soft matter science.

\newpage
\subsubsection*{\textsf{Methods summary}}
Spherical reguli (2.5\,mm in diameter, 70\,mg) were prepared by arc-melting stoichiometric mixtures of high-purity metals (Ni: 99.999; Zr: 99.97\%) in a Ti-gettered Ar-atmosphere. Mass losses during arc melting/ESL were $< 0.1$\,mg with nominal compositions checked on embedded/polished samples by EDS. ESL was performed with a custom-built levitator using an infrared laser ($P = 75$\,W, $\lambda = 810$\,nm) and an ultrafast pyrometer (Impac IGA 120-TV). 200~undercooling cycles in succession gave a statistic of nucleation events (Fig.~\ref{nucleation}). A Photron Ultima APX 775k HSC allowed the direct observation of the solidification with 10\,000\,fps and $(512 \times 320)\,$px time and space resolution. EBSD-samples were embedded in a carbon containing, electrical conductive phenolic mounting resin via application of temperature and pressure (450\,K, 25\,kN). Cylindrical resin specimens were grinded and polished (0.05\,$\mu$m colloidal alumina) with a final treatment using a slightly basic suspension (0.1\,$\mu$m colloidal silica, $p\mathrm{H}$ = 9.0). The success of the oriented embedding was examined using an optical polarization microscope (Imager.A2M, Zeiss). A LEO~1530VP SEM ($U_{\mathrm{acc}} = 20$\,kV, 1\,nm spatial resolution) equipped with an EDS-system (INCA) and an EBSD detector (HKL) was used to characterize the microstructure in terms of phase composition and crystal orientation.

\newpage
\bibliography{NiZrBib}
\bibliographystyle{plain}

\newpage
\paragraph{\textsf{Acknowledgements}}

The authors thank D.~Holland-Moritz for discussions, P.~Kuhn for comments on the manuscript and the German Science Foundation (DFG) for financial support (HE1601/28).

\newpage
\begin{figure}[!hbt]
\includegraphics[width=160mm]{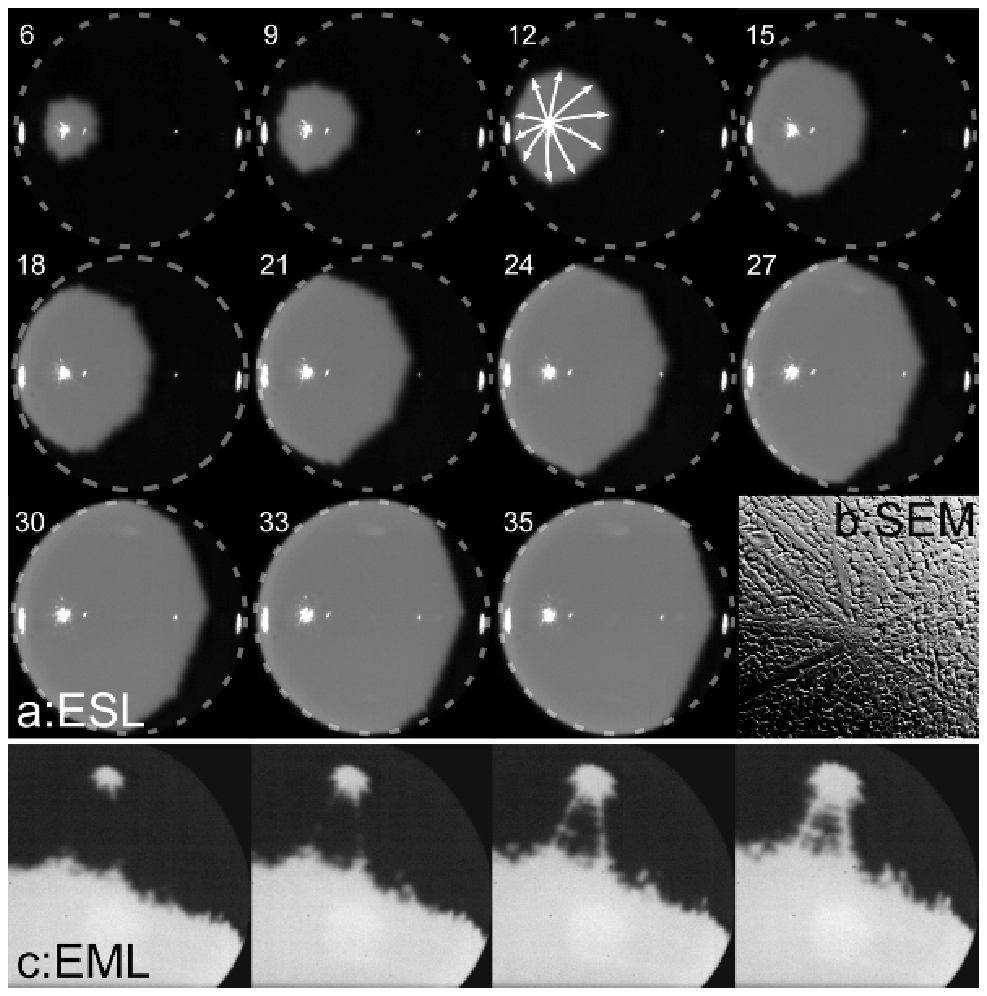}
\caption{\label{hsc} \textsf{Decagonal solidification of deeply undercooled NiZr.} HSC snapshots of decagon-shaped intersections of the moving solidification front with the surface of electrostatically (\textbf{a}) and electromagnetically (\textbf{c}) levitated NiZr droplets. Frame numbers refer to the onset point of solidification set to zero (0.1\,ms/frame). An SEM micrograph shows the geodesic surface structure (\textbf{b}). The second time series (\textbf{c}) shows the appearance of the leading dendrite tip, after traversing the sample, again with decagonal symmetry.}
\end{figure}

\newpage
\begin{figure}[!hbt]
\includegraphics[width=160mm]{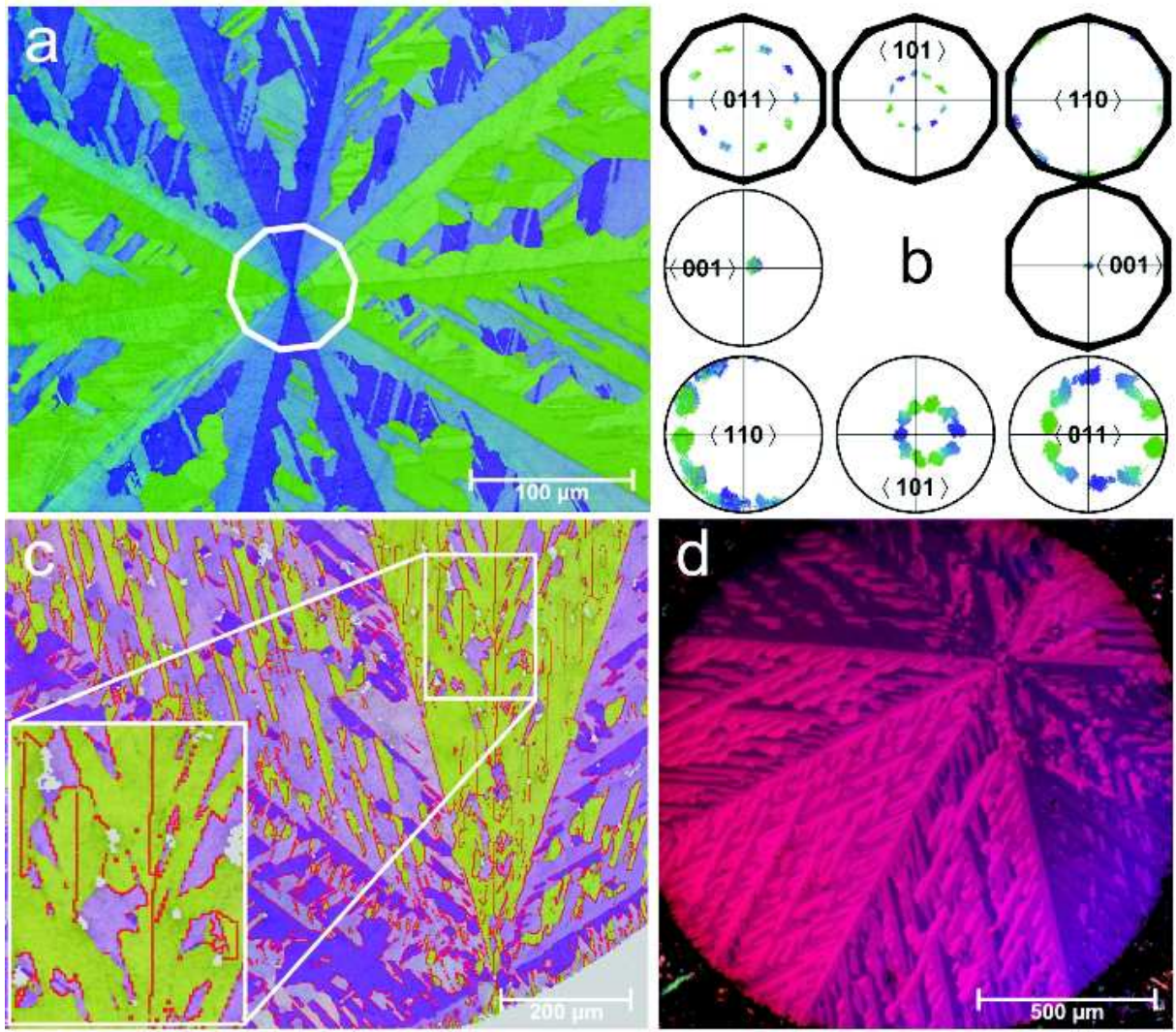}
\caption{\label{ebsd} \textsf{Tenfold twinned dendritic microstructure of NiZr.} Tenfold microtwinning apparent in the cross-sections of three distinct samples (\textbf{a},\textbf{c}: false-colour EBSD-maps; \textbf{d}: optical polarization micrograph) is confirmed by a series of IPFs (\textbf{b}): A single cluster of crystal orientations aligns with the axial growth direction along a common $[001]$-axis of the NiZr twin domains, while a perfect tenfold clustering is observed for perpendicular $\langle 110 \rangle$-directions. Circular IPFs represent the crystallographic orientations for the complete EBSD-map (\textbf{a}) of a slightly inclined sample; decagonal, alignment-corrected IPFs highlight perfect orientational relationships for the central part of the microtwin.}
\end{figure}

\newpage
\begin{figure}[!hbt]
\includegraphics[width=160mm]{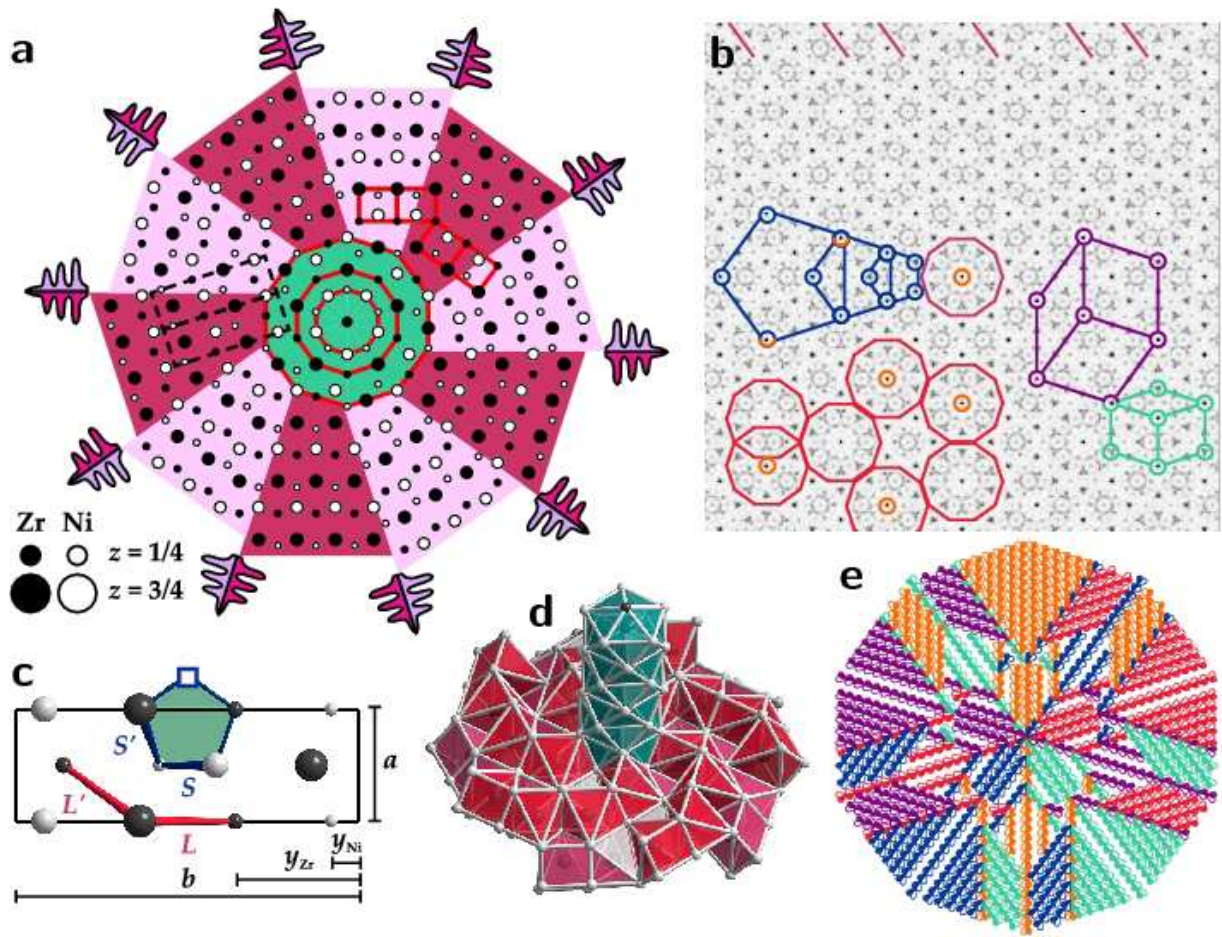}
\caption{\label{structure} \textsf{Cyclic intergrowth structure of CrB-type NiZr.} Ideal tenfold twin (\textbf{a}) and its Fourier image (\textbf{b}). Calculated Fourier amplitudes ({'}reflections{'}) exhibit several $\tau$-related scalings of regular pentagons and thin and thick Penrose rhombi. A partial tiling (covering) of decagonal clusters is visible as is a five-fold grid of parallel lines which come in two ($1\colon\tau$)-related short and long spacings. Corresponding features of decagonal long-range orientational order are present in real space (\textbf{a}), caused by defective pentagons occuring in CrB-type NiZr (\textbf{c}). Vacancies ($\square$) are attributed to geometric frustration with respect to the alternating height of atoms in the $c$-direction. A polyhedral model highlights a spiral in-plane and columnar out-of-plane growth (\textbf{d}). Twofold twinned dendrites grow into a tenfold twinned {'}snowflake{'} with its decagonal convex hull illustrating a simplified solidification front (\textbf{e}). Parallel straight and wavy twin boundaries and the polar domain structure with wedge-like junctions match the real microstructure (Fig.~\ref{ebsd}).}
\end{figure}

\newpage
\begin{figure}[!hbt]
\includegraphics[width=160mm]{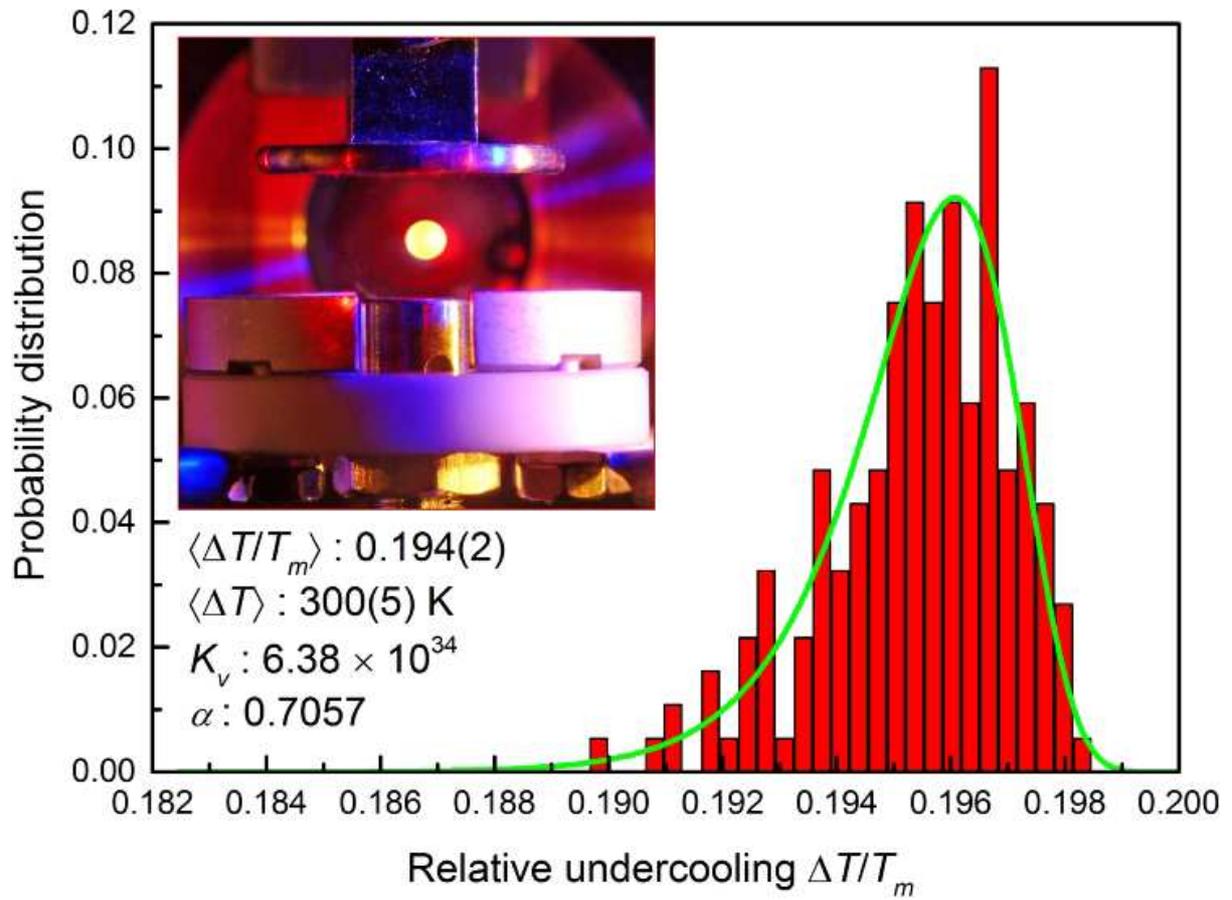}
\caption{\label{nucleation} (Extended Data Figure) \textsf{Nucleation statistics for ESL-processed NiZr.} A smooth distribution function is superimposed onto a histogram showing the normalized frequency of nucleation events versus the relative undercoolings $\Delta T/T_{\mathrm{m}}$ for 200~consecutive solidification cycles with results originating from the statistical analysis. The inset shows an ESL-processed metallic melt.}
\end{figure}

\newpage
\begin{figure}[!hbt]
\includegraphics[width=160mm]{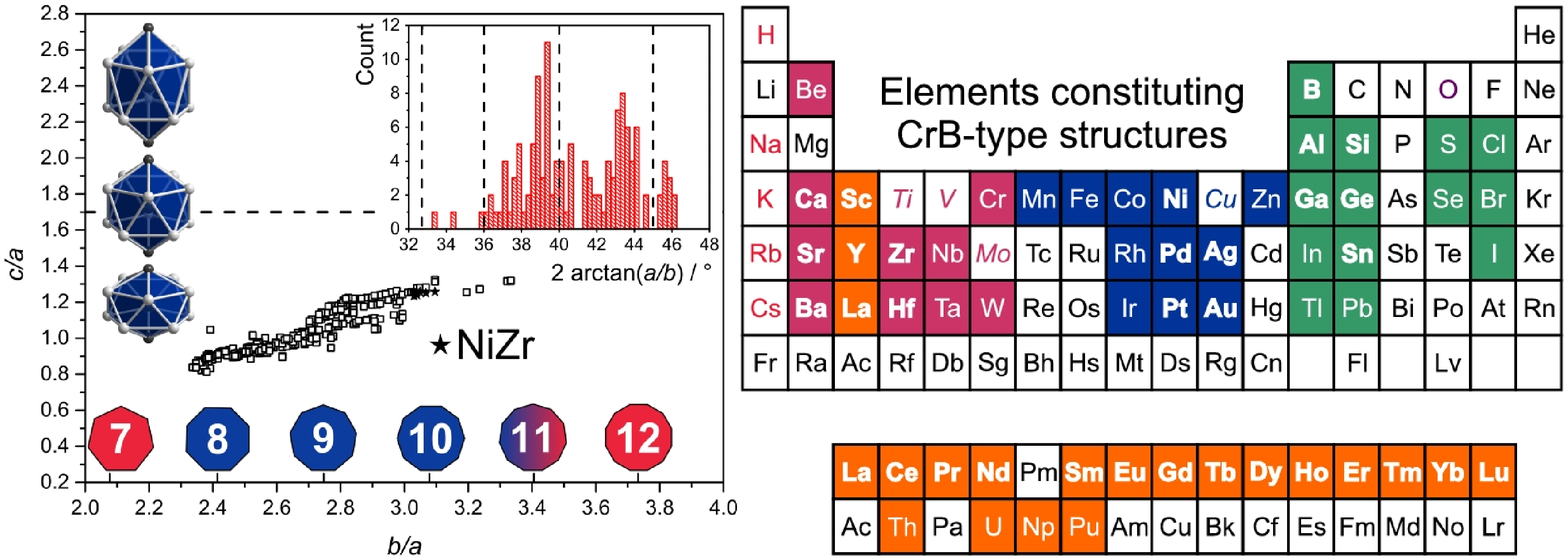}
\caption{\label{icocrb} (Extended Data Figure) \textsf{Occurrence of 312 binary CrB-type structures regarding  crystallo-geometrical constraints and element combinations.} The $b/a$ ratio determines the in-plane geometry, i.e.~the {'}polygonality{'} of the twin model. The $c/a$ ratio determines its out-of-plane geometry, i.e.~the compression/elongation of the central icosahedron, distinguishing between a (2+1)-dimensional,~decagonal and a 3D,~icosahedral case. The majority of CrB-type structures is formed between electropositive alkaline and rare earth elements and early transitions metals in combination with electronegative late transitions metals and main group elements with an additional dependance on the size ratio of the constituents. A special case is given for the alkali metals and the elements hydrogen and oxygen, since the pseudo-binary hydroxides $M$OH ($M = \mathrm{Na}, \mathrm{K}, \mathrm{Rb}, \mathrm{Cs}$ also crystallize in the CrB-type. The metals Ti, V, and Cu occur only as the minority components of ternary substitution variants of the CrB-type structure. Table~\ref{CrBtab} lists the crystallographic data for those compounds, for which the complete crystal structure, including the atomic coordinates, was determined.}
\end{figure}

\newpage
\begin{figure}[!hbt]
\includegraphics[width=160mm]{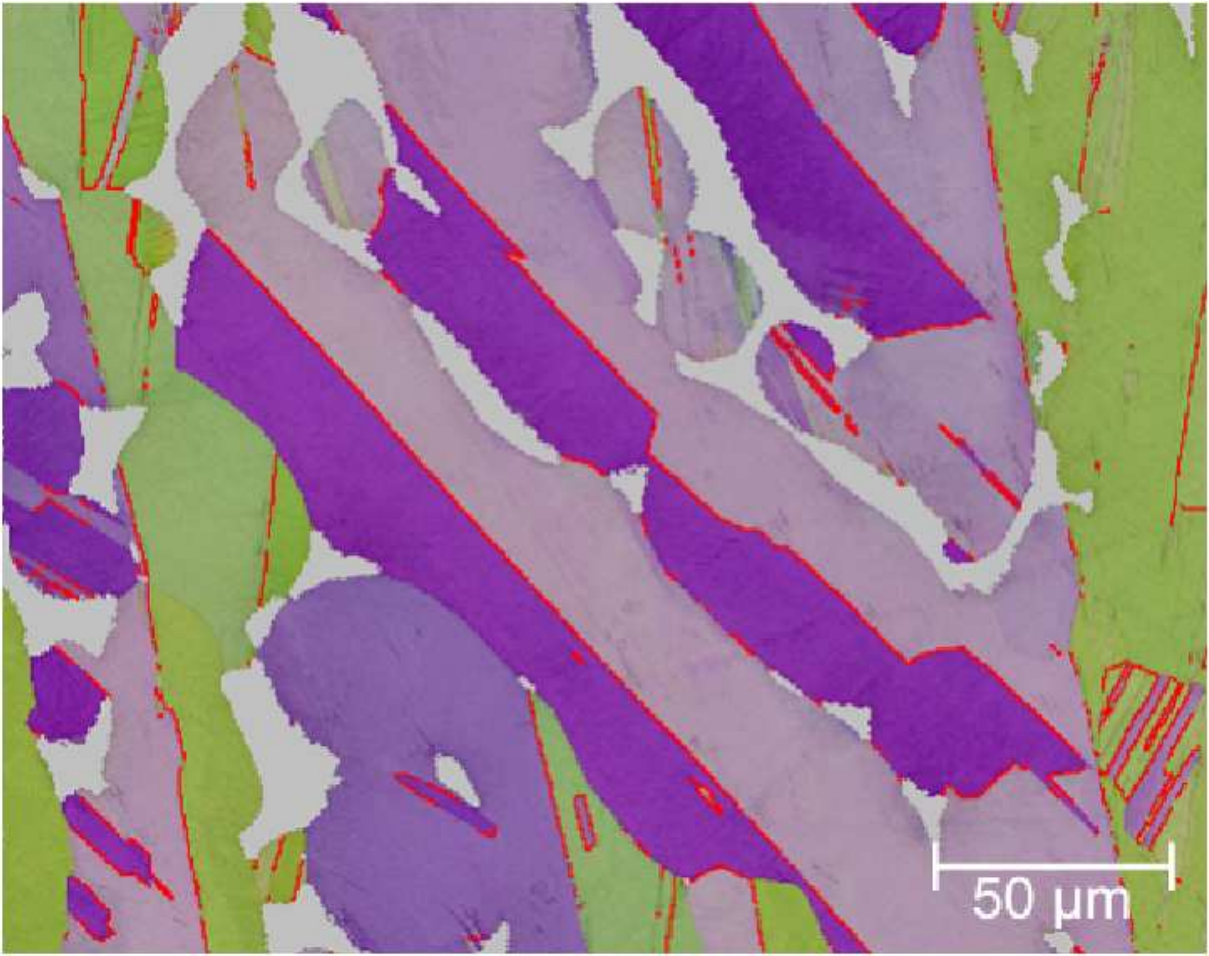}
\caption{\label{lunker} (Extended Data Figure) \textsf{Twinned dendrites as primary growth feature.} Shown are (coarsened) twinned dendrites, with their coherent (straight) 36$^{\circ}$ twin boundaries highlighted in red, surrounded by cavities (light gray areas). This proves that twinning is a genuine primary growth feature and not a result of secondary transformations affecting the sample after its solidification. Note the appearance of incoherent (wavy) twin boundaries resulting from the coalescence of neighbouring dendrites yielding an alternating pattern of parallel straight and wavy twin boundaries.}
\end{figure}

\newpage
\begin{figure}[!hbt]
\includegraphics[width=160mm]{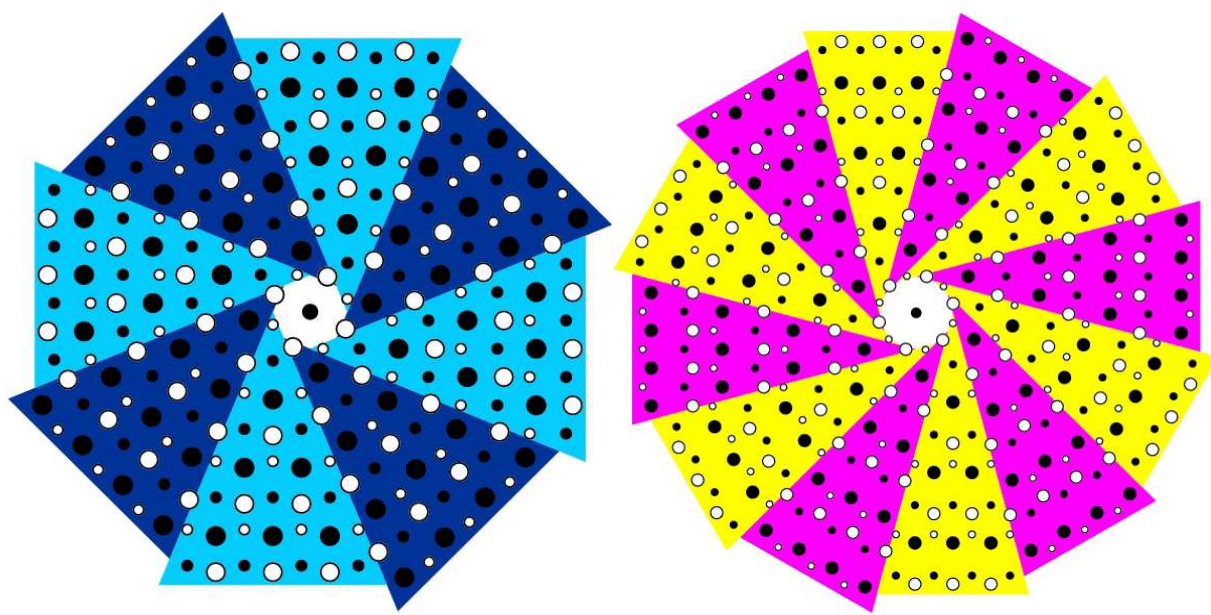}
\caption{\label{octdod} (Extended Data Figure) \textsf{Eightfold and twelvefold twinned CrB-type structure variants.} Note the opposite chirality of the spirals. See the Supplemental Information for the numerical data and general formulae.}
\end{figure}

\newpage
\begin{figure}[!hbt]
\includegraphics[width=160mm]{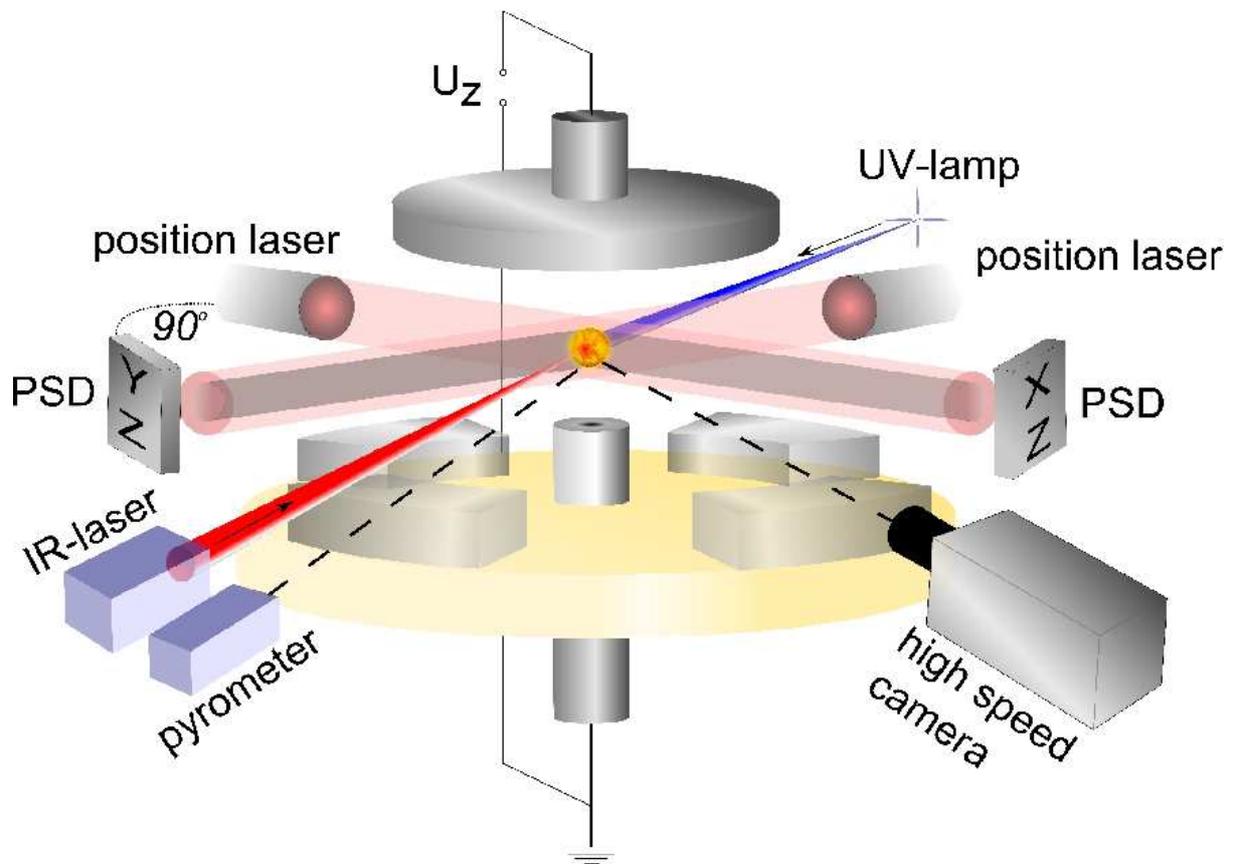}
\caption{\label{esl} (Extended Data Figure) \textsf{Electrostatic levitation.} Scheme of the experimental setup consisting of a symmetrical designed system of electrodes with a pair (quadruple) of them used for levitation and $z$-axis ($xy$-plane) positioning. A sample{'}s position is determined and adjusted in real-time by the shadow of the sample casted during orthogonal illumination with a pair of He/Ne-lasers onto oppositely mounted photosensitive detectors. An infrared laser is used for heating the sample, controlled by a contactless pyrometer. Ultraviolet radiation, via the photoelectric effect, is used to keep the surface charge balanced for leviation, prior to melting of the sample. A high-speed camera allows direct observation of solidification-related phenomena.}
\end{figure}

\newpage
\begin{figure}[!hbt]
\includegraphics[width=160mm]{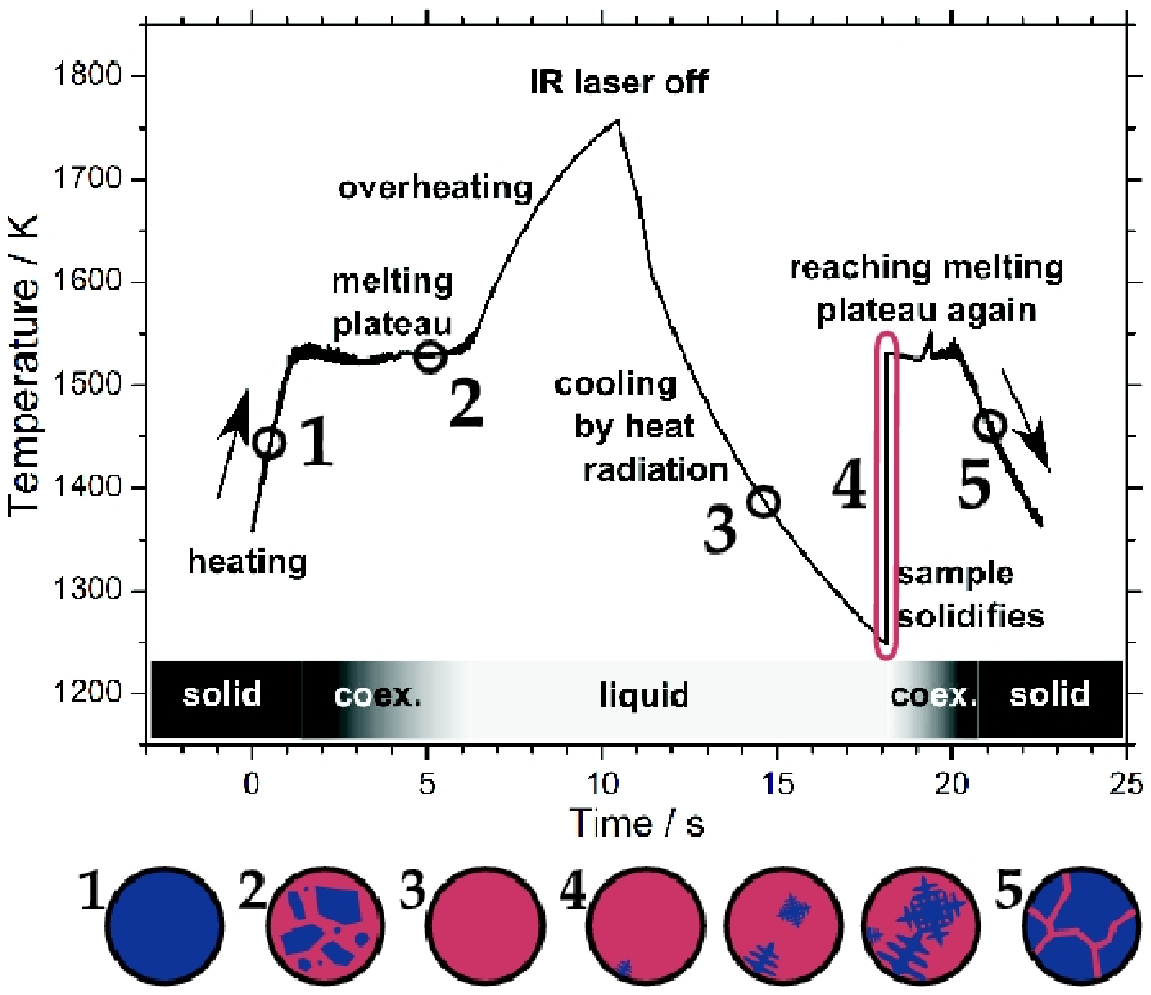}
\caption{\label{tt} (Extended Data Figure) \textsf{Temperature-time profile for NiZr solidification.} Processes  occuring inside the sample are depicted schematically.}
\end{figure}

\newpage
\begin{small}

\title{Supplemental information \\ \textsf{Quasicrystal nucleation in an intermetallic glass-former}}
\author{Wolfgang Hornfeck, Raphael Kobold, Matthias Kolbe, Dieter Herlach}

\maketitle

\newpage
\section{\textsf{Sample overview}}

In the following we give a qualitative overview over ESL-processed samples of binary NiZr, NiHf and CuZr as well as ternary (Ni,Cu)Zr and Ni(Zr,Hf) focusing on the presence ($\text{\ding{51}}$) or absence ($\text{\ding{55}}$) of features indicating tenfold twinning in either HSC-videos, optical micrographs (geodesics), EBSD-maps or IPF-plots (n.d. = not determined). Less distinctive, yet still detectable features are denoted by a bracketed symbol. 

\begin{table}[!hbt]
\begin{center}
\caption{\label{samples} \textsf{Selection of ESL-processed samples.}}
\begin{tabular}{|l|r|cccc|l|} \hline
 Sample & $\Delta T$ & \multicolumn{4}{c|}{10-fold?} & Remarks \\ 
        & (K)        & HSC & Geodesics & EBSD & IPF &         \\ \hline
 NiZr                             &  93 & \text{\ding{55}}  & \text{\ding{55}} & n.d. & n.d. & fourfold front  \\ 
 NiZr                             & 108 & \text{\ding{55}} & \text{\ding{55}} & \text{\ding{55}} & \text{\ding{51}} & fourfold front \\ 
 NiZr                             & 180 & \text{\ding{55}} & \text{\ding{51}} & \text{\ding{55}} & \text{\ding{51}} & polygonal front \\ 
 NiZr                             & 203 & \text{\ding{51}} & \text{\ding{51}} & \text{\ding{51}} & \text{\ding{51}} & antipode visible \\ 
 NiZr                             & 285 & \text{\ding{51}} & \text{\ding{51}} & \text{\ding{51}} & \text{\ding{51}} & $-$ \\ 
 NiZr                             & 298 & \text{\ding{51}} & \text{\ding{51}} & \text{\ding{55}}$^{*}$ & \text{\ding{51}} & $^{*}$preparation $\perp~[001]$ \\ \hline
 Ni$_{53.5}$Zr$_{46.5}$           & 229 & \text{\ding{55}} & \text{\ding{55}} & \text{\ding{55}} & \text{\ding{51}} & possible fourfold front \\ 
 Ni$_{49.1}$Zr$_{50.9}$           & 271 & \text{\ding{51}} & (\text{\ding{51}}) & n.d. & n.d. & vague geodesics \\ 
 Ni$_{47.6}$Zr$_{52.4}$           & 272 & \text{\ding{51}} & (\text{\ding{51}}) & (\text{\ding{51}}) & (\text{\ding{51}}) & 10-fold in small regions \\ 
 Ni$_{45.3}$Zr$_{54.7}$           & 257 & \text{\ding{55}} & \text{\ding{55}} & (\text{\ding{51}}) & (\text{\ding{51}}) & 10-fold in small regions \\ \hline
 Ni$_{50}$(Zr$_{49}$Hf)           & 291 & \text{\ding{51}} & \text{\ding{51}} & \text{\ding{51}} & \text{\ding{51}} & $-$ \\
 Ni$_{50}$(Zr$_{47.5}$Hf$_{2.5}$) & 294 & \text{\ding{51}} & \text{\ding{51}} & n.d. & n.d. & 2nd phase visible (HSC)  \\ \hline
 Ni$_{50}$(Zr$_{45}$Hf$_{5}$)     & 67 & \text{\ding{55}} & \text{\ding{55}} & \text{\ding{55}} & \text{\ding{55}} & fourfold front \\
 Ni$_{50}$(Zr$_{45}$Hf$_{5}$)     & 95 & \text{\ding{55}} & \text{\ding{55}} & n.d. & n.d. & fourfold front  \\ 
 Ni$_{50}$(Zr$_{45}$Hf$_{5}$)     & 288 & \text{\ding{51}} & \text{\ding{51}} & n.d. & n.d. & 2nd phase visible (HSC) \\ \hline
 Ni$_{50}$(Zr$_{40}$Hf$_{10}$)    & 310 & \text{\ding{55}} & (\text{\ding{51}}) & (\text{\ding{55}}) & (\text{\ding{51}}) & vague geodesics \\ \hline
 NiHf                             & 330 & \text{\ding{55}} & \text{\ding{55}} & \text{\ding{55}} & \text{\ding{51}} & fourfold front \\
 NiHf                             & 429 & \text{\ding{55}} & \text{\ding{55}} & \text{\ding{55}} & \text{\ding{51}} &  fourfold front \\ \hline
 (Ni$_{45}$Cu$_{5}$)Zr$_{50}$     & 81 & \text{\ding{55}} & \text{\ding{55}} & n.d. & n.d. & fourfold front \\ 
 (Ni$_{45}$Cu$_{5}$)Zr$_{50}$     & 306 & \text{\ding{51}} & (\text{\ding{51}}) & n.d. & n.d. & vague geodesics \\ \hline
 (Ni$_{30}$Cu$_{20}$)Zr$_{50}$    & 99 & \text{\ding{55}} & \text{\ding{55}} & n.d. & n.d. & fourfold front \\ 
 (Ni$_{30}$Cu$_{20}$)Zr$_{50}$    & 187 & \text{\ding{55}} & \text{\ding{55}} & n.d. & n.d. & fourfold front with spikes \\
 (Ni$_{30}$Cu$_{20}$)Zr$_{50}$    & 402 & \text{\ding{55}} & \text{\ding{55}} & n.d. & n.d. & spherical front with spikes \\ \hline
 (Ni$_{20}$Cu$_{30}$)Zr$_{50}$    & 99 & \text{\ding{55}} & \text{\ding{55}} & n.d. & n.d. & possible fourfold front \\ 
 (Ni$_{20}$Cu$_{30}$)Zr$_{50}$    & 174 & \text{\ding{55}} & \text{\ding{55}} & n.d. & n.d. & $-$ \\ 
 (Ni$_{20}$Cu$_{30}$)Zr$_{50}$    & 377 & \text{\ding{55}} & \text{\ding{55}} & n.d. & n.d. & $-$  \\  \hline
 (Ni$_{5}$Cu$_{45}$)Zr$_{50}$     & 65 & \text{\ding{55}} & \text{\ding{55}} & n.d. & n.d. & spherical front  \\
 (Ni$_{5}$Cu$_{45}$)Zr$_{50}$     & 346 & \text{\ding{55}} & \text{\ding{55}} & n.d. & n.d. & spherical front  \\ \hline
  CuZr                            & 104 & \text{\ding{55}} & \text{\ding{55}} & n.d. & n.d. & spherical front \\
	CuZr                            & 297 & \text{\ding{55}} & \text{\ding{55}} & n.d. & n.d. & spherical front \\ \hline
\end{tabular}
\end{center}
\end{table}

\newpage

\section{\textsf{Construction of the twin model}}

\subsection{\textsf{General formulae}}

For a compound AB crystallizing in the orthorhombic CrB-type structure with lattice parameters $a$, $b$, $c$ we define a pair of distances
\begin{equation}
S = \sqrt{d_{\mathrm{AA}}^{2}-(c/2)^2} \qquad \mathrm{and} \qquad
L = \sqrt{d_{\mathrm{BB}}^{2}-(c/2)^2}\,,
\end{equation}
where $d_{\mathrm{AA}}$ and $d_{\mathrm{BB}}$ are the shortest interatomic distances between alike constituents (here, $\mathrm{A} = \mathrm{Ni}$ and $\mathrm{B} = \mathrm{Zr}$). In fact, $S$ and $L$ represent \emph{projections} -- along the $c$-direction -- of the distances $d_{\mathrm{AA}}$ and $d_{\mathrm{BB}}$, respectively. Another pair of distances is given by
\begin{equation}
 S{'} = \sqrt{\left(\frac{a}{2}\right)^{2}+\left(\frac{L}{2}-\frac{S}{2}\right)^{2}} \qquad \mathrm{and} \qquad
 L{'} = \sqrt{\left(\frac{a}{2}\right)^{2} + \left(\frac{b}{2}-L\right)^{2}}\,,
\end{equation}
as shown in Fig.~3c. For the construction of the ideal $n$-fold twinned structure model we impose the restrictions
\begin{equation}
 S{'} \stackrel{!}{=} S \Rightarrow \mathrm{equation~(\ref{eqS})} \qquad \mathrm{and} \qquad
 L{'} \stackrel{!}{=} L \Rightarrow \mathrm{equation~(\ref{eqL})}\,,
\end{equation}
which allow to derive a complete and consistent set of structural parameters via elementary geometrical relations:
\begin{eqnarray}
\varphi/2 & = & \frac{360^{\circ}}{2\,n}\,, \label{eqphi} \\
b & = & \frac{1}{\tan\varphi/2}\,a\,, \label{eqa} \\
L & = & \frac{a^{2}+b^{2}}{4\,b}\,, \label{eqL} \\
y_{\mathrm{B}} & = & \frac{b-L}{2\,b}\,, \\
S & = & \frac{1}{3}\left(-L + \sqrt{3\,a^{2} + 4\,L^{2}}\right)\,, \label{eqS} \\
y_{\mathrm{A}} & = & \frac{S}{2\,b}\,, \mathrm{~and} \label{eqe} \\
\sigma & = & \frac{a}{2} = \frac{d}{2\,\sqrt{1+(b/a)^{2}}}\,.\label{eqsigma}
\end{eqnarray}
This includes special values for the atomic coordinates $y_{\mathrm{A}}$ and $y_{\mathrm{B}}$ of the Wyckoff position $4c~(0,y,1/4)$ for both constituents. Note, that all parameters given in succession solely depend on the number $n$ of twin domains and refer to a common arbitrary scaling factor given by the lattice parameter $a$ (which may be set to unity). Note also, that for the special case of $n = 10$ (NiZr) the geometrical match is perfect, such that the crystal structure across the twin boundaries is identical to the crystal structure of bulk NiZr.

\subsection{\textsf{Specific cases}}

The following table lists the calculated values for the parameters $\varphi$, $b$, $L$, $y_{\mathrm{B}}$, $S$, $y_{\mathrm{A}}$, and $\sigma$ defined in Equations~\ref{eqphi} to~\ref{eqsigma} for different choices of $n = 8, 10, 12$, with $a$ set to unity, in analytical and numerical form, and in comparison to NiZr (compare Fig.~\ref{octdod}). 

\begin{table}[!hbt]
\begin{center}
\begin{small}
\caption{\label{cases} \textsf{Ideal geometric parameters for $n$-fold twin structures ($n = 8,10,12$).}}
\begin{tabular}{|c|cc|l|} \hline
         & \multicolumn{2}{c|}{$n=8$} &  \\ \hline
 $\varphi/2$                                                 & & 22.5$^{\circ}$ & \\
 $b/a$     & $1+\sqrt{2}$                                     & 2.414214 &  \\
 $L/a$     & $1/\sqrt{2}$                                     & 0.707107 &  \\
 $y_{\mathrm{B}}/b$ & $1/(2\sqrt{2})$                                  & 0.353553 &  \\
 $S/a$     & $(-1+\sqrt{10})/(3\sqrt{2})$                     & 0.509654 &  \\
 $y_{\mathrm{A}}/b$ & $-(1/12) (-1 + \sqrt{2}) (\sqrt{2} - 2\sqrt{5})$ & 0.105553 &  \\ 
 $\sigma/d$ & $1/(2\sqrt{2(2+\sqrt{2})})$ & 0.191342 &  \\ \hline
         & \multicolumn{2}{c|}{$n=10$} & \multicolumn{1}{c|}{NiZr} \\ \hline
 $\varphi/2$                                                 & & 18$^{\circ}$ & 18.2$^{\circ}$ \\        
 $b/a$     & $\sqrt{5+2\sqrt{5}}$         & 3.077684 & 3.040698 \\
 $L/a$     & $\sqrt{(1/10)(5+\sqrt{5})}$  & 0.850651 & 0.868579 \\
 $y_{\mathrm{B}}/b$ & $(1/20)(5+\sqrt{5})$         & 0.361803 & 0.3609(8) \\
 $S/a$     & $\sqrt{(1/10)(5-\sqrt{5})}$  & 0.525731 & 0.534687 \\
 $y_{\mathrm{A}}/b$ & $(1/20)(-5+3\sqrt{5})$       & 0.085410 & 0.0817(17) \\ 
 $\sigma/d$ & $(1/8)(-1+\sqrt{5})$ & 0.154508 &  \\ \hline 
         & \multicolumn{2}{c|}{$n=12$} &  \\ \hline
 $\varphi/2$                                                 & & 15$^{\circ}$ & \\         
 $b/a$     & $2+\sqrt{3}$                             & 3.732051 &  \\
 $L/a$     & $1$                                      & 1.000000 &  \\
 $y_{\mathrm{B}}/b$ & $(1/2)(-1+\sqrt{3})$                     & 0.366025 &  \\
 $S/a$     & $(1/3)(-1+\sqrt{7})$                     & 0.548584 &  \\
 $y_{\mathrm{A}}/b$ & $-(1/6) (-2 + \sqrt{3}) (-1 + \sqrt{7})$ & 0.073496 &  \\ 
 $\sigma/d$ & $1/(4\sqrt{2+\sqrt{3}})$ & 0.129410 &  \\ \hline 
\end{tabular}
\end{small}
\end{center}
\end{table}

\newpage
\section{\label{nuclskrip} \textsf{Details regarding homogeneous nucleation}}

\subsection{\textsf{Skripov analysis}}

In order to distinguish between heterogeneous or homogeneous nucleation taking place in an undercooled melt, a statistical analysis of nucleation events can be performed, if a single sample is repeatedly molten, undercooled and solidified with its relative undercoolings $\Delta T/T_{\mathrm{m}}$ determined. A histogram mapping the (normalized) frequency of nucleation events versus $\Delta T/T_{\mathrm{m}}$ (Fig.~\ref{nucleation}) allows the extraction of key parameters of nucleation and eventually the distinction between the heterogeneous and the homogeneous case.

For this purpose it is necessary to describe the distribution function of nucleation events by some physical model. Following Skripov{'}s$^{(a)}$ assumption of Poisson-distributed nucleation events a probability distribution 
\begin{equation}
\omega(T) = \frac{I_{\mathrm{ss}} V}{\dot{T}} \exp
\left(- \int_{T_{\mathrm{L}}}^{T} \frac{I_{\mathrm{ss}} V}{\dot{T}} dT\right)
\end{equation}
is derived ($\dot{T}$ denoting the time derivative of the temperature). Here, $I_{\mathrm{ss}}$ is the unknown nucleation rate, which can be modelled, using the approximation$^{(b)}$ $\Delta G_{V} = \Delta S_{\mathrm{f}} \, \Delta T \, V_{\mathrm{m}}^{-1}$, as 
 \begin{equation}
I_{\mathrm{ss}} = K_{V} \cdot \exp\left(-\frac{\Delta G ^{*}}{k_{\mathrm{B}}T_{\mathrm{n}}}\right)\,,
\end{equation}
with $T_{\mathrm{n}}$ the nucleation temperature and the pre-exponential factor 
\begin{equation}
K_{V} = \frac{k_{\mathrm{B}}T_{\mathrm{n}}}{3\,a_{0}^{3}\,\eta(T)}N_{0}\,
\end{equation}
of the nucleation rate. Here, $a_{0}$ denotes a typical (nearest-neighbour) interatomic distance, $\eta(T)$ is the temperature-dependent dynamical viscosity and $N_{0}$ gives the number of nuclei, with $N_{0} = N_{\mathrm{A}}/V_{\mathrm{m}}$ in the limit, in which all $N_{\mathrm{A}}$ atoms within the molar volume $V_{\mathrm{m}}$ act as nuclei. $K_{V}$ thus represents a sensitive measure for the degree of homogeneous nucleation.

Now, the cumulative distribution function
\begin{equation}
F(T) = 1 - \exp\left(- \frac{V}{\dot T} \int_{T_{\mathrm{L}}}^{T} I_{\mathrm{ss}} dT \right)
\end{equation}
resulting from the integration of $\omega(T)$, is simplified by neglecting the temperature dependence of $K_{\mathrm{V}}$ and written as
\begin{equation}
F(T) = 1 - \exp\left(- \frac{V K_{\mathrm{V}}}{\dot T \displaystyle\frac{d \left(\Delta G^{*}/(k_{\mathrm{B}}\,T)\right)}{dT}} \int_{T_{\mathrm{L}}}^{T} \left(\frac{CT^2}{\Delta T^2}\right) \right)
\end{equation}
with $-C\,T^2/\Delta T^2 = -\Delta G ^{*}/k_{\mathrm{B}}T_{\mathrm{n}}$.

Plotting $\ln(-\ln(1-F(T)))$ versus $T^2/\Delta T^2$ leads to a linear relation from which $K_{\mathrm{V}}$ and $\Delta G^{*}$ are determined (compare Tab.~\ref{nuclstat}).

\subsection{\textsf{Calculation of the critical radius}}

According to classical nucleation theory (CNT) the critical radius $r^{*}$ is given as
\begin{equation}
r^{*} = \frac{2\,\sigma}{\Delta G_{\mathrm{sl}}}
\end{equation} 
where 
\begin{equation}
\Delta G_{\mathrm{sl}} = G_{\mathrm{s}} - G_{\mathrm{l}} = \sqrt{\frac{16\,\pi\,\sigma^{3}}{3\,\Delta G^{*}}}
 \end{equation}
denotes the difference in the free enthalpy between the solid and the liquid phase. Thus, $r^{*}$ can be calculated, if the solid-liquid interfacial energy $\sigma$ and the critical free enthalpy of nucleation $\Delta G^{*}$ are known. While $\Delta G^{*}$ is obtained in the aforementioned way, the  solid-liquid interfacial energy $\sigma$ is calculated based on the negentropic model of Spaepen$^{c}$ according to
\begin{equation}
\sigma = \alpha \frac{\Delta H_{\mathrm{cr}}}{\left(N_{\mathrm{A}}^{\phantom{2}}V_{\mathrm{m}}^{2}\right)^{1/3}}\,.
\end{equation}
Here, $\Delta H_{\mathrm{cr}}$ denotes the enthalpy of crystallization and $\alpha$ is the dimensionless interfacial energy. Table~\ref{nuclstat} lists the parameters used for and obtained from the statistical evaluation of a series of nucleation events.

\begin{table}[!hbt]
\caption{\label{nuclstat} \textsf{Parameters of NiZr used in the evaluation of the nucleation statistics (top) or resulting from it (bottom).}}
\begin{tabular}{lllll} \hline
 Parameter & Symbol & Value & Unit & Ref. \\ \hline
 Liquidus temperature & $T_{L}$ & 1533 & K & (d) \\
 Enthalpy of crystallization & $\Delta H_{\mathrm{cr}}$ & $7.63(4)$ & kJ/mol & (e) \\
 Entropy of crystallization & $\Delta S_{\mathrm{cr}} = \Delta H_{\mathrm{cr}}/T_{L}$ & $5.07405$ & J/(mol\,K) & calc. \\
 Cooling rate & $\Delta T/\Delta t$ & 34 & K/s & meas. \\
 Sample radius & $r$ & $1.29 \times 10^{-3}$ & m & meas. \\
 Sample volume & $V = (4/3)\pi\,r^{3}$ & $8.95 \times 10^{-9}$ & $\mathrm{m}^{3}$ & meas. \\
 Molar volume NiZr & $\left(V_{\mathrm{m}} = N_{\mathrm{A}}V_{\mathrm{cr}}\big/Z\right)$ & $1.03 \times 10^{-5}$ & $\mathrm{m}^{3}/\mathrm{mol}$ & calc. \\ 
 & $\left(V_{\mathrm{m, Ni}}+V_{\mathrm{m, Zr}}\right)/2$ & & & \\ \hline
 Critical free enthalpy & $\Delta G^{*}$ & 59.717 & $k_\mathrm{B} T_{\mathrm{n}}$ & see~\ref{nuclskrip} \\ 
 Solid-liquid interfacial energy & $\sigma$ & 0.1346(7) & $\mathrm{J}/\mathrm{m}^{2}$ &  \\
 Dimensionless interfacial energy & $\alpha$ & 0.7057 & $-$ &  \\ 
 Pre-exponential factor & $K_{V}$ & $6.38 \times 10^{34}$ & $\mathrm{m}^{-3}\mathrm{s}^{-1}$ &  \\
 Critical radius & $r^{*}$ & 1.3 & nm &  \\ \hline
\end{tabular}
\end{table}~\begin{tabular}{ll}
 (a) & Skripov V. P., Baidakov, V. G. \& Kaverin, A. M. (1979). \textit{Physica A} \textbf{95}, 169--180.\\
 (b) & Turnbull, D. (1950). \textit{J. Appl. Phys.} \textbf{21}, 1022--1028.\\
 (c) & Spaepen, F. (1975). \textit{Acta Met.} \textbf{23}, 729--743.\\
 (d) & Nash, P. \& Jayanth, C. S. (1984). \textit{Bull. Alloy Phase Diagr.} \textbf{5}, 144-- 148.\\
 (e) & Henaff, M. P., Colinet, C., Pasturel, A. \& Buschow, K. H. J. (1984). \textit{J. Appl. Phys.} \textbf{56}, 307--310. 
\end{tabular}

\newpage
\section{\textsf{Occurence of CrB-type structures}}

The following table lists a selection of crystallographic data for those 132~binary CrB-type structures for which the complete crystal structure has been determined.

\begin{table}[!hbt]
\begin{center}
\begin{footnotesize}
\caption{\label{CrBtab} \textsf{Crystallographic data of binary CrB-type structures.}}
\begin{tabular}{lcccccll}
Formula	&	$a$/pm	&	$b$/pm	&	$c$/pm	&	$a/b$	&	$c/a$	&	$y_A$	&	$y_B$ \\ \hline	
AgCa	&	405.22 &	1144.70	&	464.43 & 0.354 & 1.146 &	0.0741(3)	&	0.35773(10)	\\
AgCl	&	332.00 &	\phantom{3}983.50	&	410.80 & 0.338 & 1.237 &	0.102(1)	&	0.359(1)	\\
AlHf	&	325.30 &	1082.20	&	428.00 & 0.301 & 1.316 &	0.108	&	0.367	\\
AlTh	&	442.00 &	1145.00	&	419.00 & 0.386 & 0.948 &	0.057	&	0.353	\\
AlY	  &	388.40 &	1152.20	&	438.50 & 0.337 & 1.129 &	0.07	&	0.35	\\
AlZr	&	335.30 &	1086.60	&	426.60 & 0.309 & 1.272 &	0.076	&	0.334	\\
AuCa	&	396.10 &	1107.50	&	457.60 & 0.358 & 1.155 &	0.08	&	0.36	\\
AuCe	&	390.00 &	1114.00	&	475.00 & 0.350 & 1.218 &	0.065	&	0.365	\\
AuDy	&	371.00 &	1087.00	&	461.00 & 0.341 & 1.243 &	0.065	&	0.365	\\
AuEr	&	365.00 &	1081.00	&	458.00 & 0.338 & 1.255 &	0.065	&	0.365	\\
AuGd	&	376.00 &	1094.00	&	464.00 & 0.344 & 1.234 &	0.065	&	0.365	\\
AuHo	&	368.50 &	1083.50	&	459.50 & 0.340 & 1.247 &	0.065	&	0.365	\\
AuLa	&	395.00 &	1120.00	&	478.00 & 0.353 & 1.210 &	0.065	&	0.365	\\
AuNd	&	384.00 &	1107.00	&	470.00 & 0.347 & 1.224 &	0.065	&	0.365	\\
AuPr	&	387.00 &	1110.00	&	472.00 & 0.349 & 1.220 &	0.065	&	0.365	\\
AuSm	&	380.00 &	1100.00	&	466.00 & 0.345 & 1.226 &	0.065	&	0.365	\\
AuTb	&	373.00 &	1090.00	&	462.00 & 0.342 & 1.239 &	0.065	&	0.365	\\
AuTm	&	362.00 &	1078.00	&	457.00 & 0.336 & 1.262 &	0.065	&	0.365	\\
BCr  	&	297.82 &	\phantom{3}787.00	&	293.46 & 0.378 & 0.985 &	0.064(2)	&	0.35475(3)	\\
BTa  	&	327.60 &	\phantom{3}866.90	&	315.70 & 0.378 & 0.964 &	0.06	&	0.354	\\
CoTh	&	374.00 &	1088.00	&	416.00 & 0.344 & 1.112 &	0.084	&	0.364	\\
CoY  	&	410.60 &	1035.80	&	390.60 & 0.396 & 0.951 &	0.068(2)	&	0.354(1)	\\
GaCa	&	419.30 &	1144.70	&	438.47 & 0.366 & 1.046 &	0.06651(6)	&	0.35401(11)	\\
GaCe	&	447.30 &	1133.90	&	420.10 & 0.394 & 0.939 &	0.08	&	0.361	\\
GaDy	&	430.00 &	1089.00	&	406.70 & 0.395 & 0.946 &	0.074	&	0.36	\\
GaGd	&	433.88 &	1101.04	&	410.58 & 0.394 & 0.946 &	0.0717(1)	&	0.3583(1)	\\
GaHo	&	428.10 &	1077.40	&	405.00 & 0.397 & 0.946 &	0.0763	&	0.3579	\\
GaLa	&	449.30 &	1134.10	&	420.80 & 0.396 & 0.937 &	0.083	&	0.362	\\
GaNd	&	439.00 &	1134.00	&	419.70 & 0.387 & 0.956 &	0.082	&	0.378	\\
GaPr	&	446.00 &	1133.80	&	419.90 & 0.393 & 0.941 &	0.08	&	0.359	\\
GaSc	&	402.20 &	1020.50	&	389.50 & 0.394 & 0.968 &	0.083	&	0.362	\\
GaSm	&	433.00 &	1134.00	&	419.60 & 0.382 & 0.969 &	0.079	&	0.378	\\
GaTb	&	433.00 &	1090.00	&	409.00 & 0.397 & 0.945 &	0.074	&	0.359	\\
GaY  	&	430.20 &	1086.00	&	407.30 & 0.396 & 0.947 &	0.08300	&	0.362	\\ \hline
\end{tabular}
\end{footnotesize}
\end{center}
\end{table}

\begin{table}[!hbt]
\begin{center}
\begin{footnotesize}
\caption{\textsf{Crystallographic data of binary CrB-type structures (continued).}}
\begin{tabular}{lcccccll}
Formula	&	$a$/pm	&	$b$/pm	&	$c$/pm	&	$a/b$	&	$c/a$	&	$y_A$	&	$y_B$ \\ \hline	
GeBa	&	505.70 &	1194.20	&	429.90 & 0.423 & 0.850 &	0.064(3)	&	0.36(4)	\\
GeDy	&	425.40 &	1062.30	&	390.40 & 0.400 & 0.918 &	0.0773(4)	&	0.3615(7)	\\
GeEr	&	421.99 &	1058.10	&	390.60 & 0.399 & 0.926 &	0.0859(8)	&	0.3613(8)	\\
GeEu	&	471.50 &	1126.00	&	410.10 & 0.419 & 0.870 &	0.071	&	0.361	\\
GeGd	&	433.90 &	1078.80	&	397.30 & 0.402 & 0.916 &	0.08	&	0.36	\\
GeNd	&	447.80 &	1107.50	&	404.30 & 0.404 & 0.903 &	0.0798(4)	&	0.3611(4)	\\
GePr	&	447.20 &	1107.00	&	402.90 & 0.404 & 0.901 &	0.083(1)	&	0.359(1)	\\
GeS	  &	365.76 &  \phantom{3}956.88	&	327.23 & 0.382 & 0.895 &	0.11062	&	0.3581	\\
GeSc	&	400.70 &	1006.00	&	376.20 & 0.398 & 0.939 &	0.083	&	0.362	\\
GeSm	&	438.70 &	1089.00	&	399.30 & 0.403 & 0.910 &	0.08	&	0.36	\\
GeSr	&	480.80 &	1136.00	&	416.90 & 0.423 & 0.867 &	0.07(2)	&	0.362(3)	\\
GeTb	&	428.26 &	1068.02	&	392.90 & 0.401 & 0.917 &	0.0844(5)	&	0.3611(5)	\\
GeTm	&	418.50 &	1052.40	&	388.50 & 0.398 & 0.928 &	0.06	&	0.354	\\
GeY  	&	426.20 &	1069.40	&	394.10 & 0.399 & 0.925 &	0.083	&	0.362	\\ 
HoGe	&	424.28 &	1062.30	&	391.90 & 0.399 & 0.924 &	0.0832(6)	&	0.3619(6)	\\
InBr	&	461.31 &	1271.20	&	469.96 & 0.363 & 1.019 &	0.111	&	0.3508	\\
InCl	&	424.20 &	1232.00	&	468.90 & 0.344 & 1.105 &	0.11189(31)	&	0.34525(8)	\\
InI	  &	451.82 &	1149.20	&	418.61 & 0.393 & 0.927 &	0.094	&	0.3567	\\
NiCe	&	379.40 &	1054.60	&	436.70 & 0.360 & 1.151 &	0.0763	&	0.36134	\\
NiGd	&	377.13 &	1032.72	&	424.88 & 0.365 & 1.127 &	0.0789(2)	&	0.3571(5)	\\
NiNd	&	380.59 &	1046.20	&	433.45 & 0.364 & 1.139 &	0.0727(5)	&	0.36149(1)	\\
NiZr 	&	327.12 &	\phantom{3}993.10	&	410.72 & 0.329 & 1.256 &	0.084(0)	&	0.36(0)	\\
PbBa	&	529.00 &	1260.00	&	478.00 & 0.420 & 0.904 &	0.0802(3)	&	0.3734(8)	\\
PdGd	&	373.60 &	1055.00	&	454.80 & 0.354 & 1.217 &	0.085	&	0.36	\\
PbSr	&	501.80 &	1223.00	&	464.80 & 0.410 & 0.926 &	0.078	&	0.368	\\
PdZr	&	333.05 &	1030.40	&	437.45 & 0.323 & 1.313 &	0.0876(2)	&	0.3572(2)	\\
PrNi	&	383.07 &	1054.30	&	436.90 & 0.363 & 1.141 &	0.0724(1)	&	0.36178(4)	\\
PtCe	&	388.40 &	1084.50	&	451.70 & 0.358 & 1.163 &	0.0872(7)	&	0.3638(10)	\\
PtNd	&	384.60 &	1076.90	&	454.20 & 0.357 & 1.181 &	0.084	&	0.369	\\
PtPr	&	389.10 &	1089.90	&	456.90 & 0.357 & 1.174 &	0.091	&	0.368	\\
PtTh	&	390.00 &	1109.00	&	445.40 & 0.352 & 1.142 &	0.09(2)	&	0.36(4)	\\
PtU  	&	370.30 &	1079.20	&	439.10 & 0.343 & 1.186 &	0.0889(3)	&	0.362(2)	\\
PtZr	&	340.82 &	1030.00	&	428.06 & 0.331 & 1.256 &	0.0904(1)	&	0.3582(1)	\\ 
PuNi	&	359.00 &	1021.00	&	422.00	&	0.352	&	1.175	&	0.0779(9)	&	0.3579(34)	\\
RhTh	&	386.60 &	1124.00	&	422.00	&	0.344	&	1.092	&	0.09(2)	&	0.36(4)	\\
SiBa	&	495.10 &	1178.70	&	401.90	&	0.420	&	0.812	&	0.058	&	0.358	\\
SiCa	&	451.92 &	1067.80	&	387.34	&	0.423	&	0.857	&	0.0697	&	0.3616	\\
SiDy	&	424.72 &	1047.55	&	380.08	&	0.405	&	0.895	&	0.0669(7)	&	0.3586(2)	\\
SiEr	&	419.50 &	1035.30	&	377.90	&	0.405	&	0.901	&	0.087(1)	&	0.358(2)	\\
SiEu	&	469.58 &	1112.42	&	397.99	&	0.422	&	0.848	&	0.0649(8)	&	0.36(2)	\\
SiHo	&	423.13 &	1044.74	&	380.54	&	0.405	&	0.899	&	0.077(1)	&	0.3591(1)	\\ 
SiSc	&	398.80 &	\phantom{3}988.20	&	365.90	&	0.404	&	0.918	&	0.081(1)	&	0.36(1)	\\
SiSr	&	482.60 &	1133.70	&	405.20	&	0.426	&	0.840	&	0.0639	&	0.361	\\
SiY  	&	425.10 &	1052.60	&	382.60	&	0.404	&	0.900	&	0.06	&	0.354	\\
SiYb	&	417.80 &	1031.00	&	376.80	&	0.405	&	0.902	&	0.075	&	0.361	\\
SnBa	&	531.00 &	1248.50	&	465.00	&	0.425	&	0.876	&	0.075	&	0.368	\\
SnCa	&	476.61 &	1153.80	&	434.52	&	0.413	&	0.912	&	0.0857	&	0.3669	\\
SnEu	&	497.60 &	1190.00	&	445.60	&	0.418	&	0.896	&	0.076	&	0.365	\\
SnS  	&	413.60 &	1148.80	&	417.20	&	0.360	&	1.009	&	0.1251(1)	&	0.35(2)	\\
SnSe	&	430.70 &	1171.60	&	430.70	&	0.368	&	1.000	&	0.124(1)	&	0.356(2)	\\
SnSr	&	504.50 &	1204.00	&	449.40	&	0.419	&	0.891	&	0.0788(4)	&	0.3633(3)	\\ 
TlI	  &	457.73 &	1278.30	&	516.35	&	0.358	&	1.128	&	0.1079	&	0.368	\\
ZnCa	&	420.20 &	1161.00	&	444.20	&	0.362	&	1.057	&	0.065	&	0.355	\\ \hline
\end{tabular}
\end{footnotesize}
\end{center}
\end{table}

\end{small}

\end{onecolumn}
\end{document}